\documentclass[]{article}
\makeatletter\if@twocolumn\PassOptionsToPackage{switch}{lineno}\else\fi\makeatother

\usepackage{amsfonts,amssymb,amsbsy,latexsym,amsmath,tabulary,graphicx,times,caption,fancyhdr,setspace}
\usepackage[utf8]{inputenc}
\usepackage{graphicx}
\usepackage{amsmath}
\usepackage{subcaption}
\usepackage{booktabs}
\usepackage[bottom]{footmisc}
\graphicspath{{picture/}}
\usepackage{lscape}
\usepackage{tablefootnote}
\usepackage[nameinlink,capitalise]{cleveref}
\Crefname{figure}{Fig.}{Figs.}

\usepackage{url,multirow,morefloats,floatflt,cancel,tfrupee}
\makeatletter
\AtBeginDocument{\@ifpackageloaded{textcomp}{}{\usepackage{textcomp}}}
\makeatother
\usepackage{colortbl}
\usepackage{xcolor}
\usepackage{pifont}
\usepackage[nointegrals]{wasysym}
\usepackage[justification=centering]{caption}
\usepackage{hyperref}
\hypersetup{
    colorlinks = true,
    citecolor = {blue}
}
\makeatletter

\def\mcWidth#1{\csname TY@F#1\endcsname+\tabcolsep}

\def\cAlignHack{\rightskip\@flushglue\leftskip\@flushglue\parindent\z@\parfillskip\z@skip}
\def\rAlignHack{\rightskip\z@skip\leftskip\@flushglue \parindent\z@\parfillskip\z@skip}

\@ifundefined{etal}{}{}

\usepackage{ifxetex}
\ifxetex\else\if@twocolumn\@ifpackageloaded{stfloats}{}{\usepackage{dblfloatfix}}\fi\fi

\AtBeginDocument{
\expandafter\ifx\csname eqalign\endcsname\relax
\def\eqalign#1{\null\vcenter{\def\\{\cr}\openup\jot\m@th
  \ialign{\strut$\displaystyle{##}$\hfil&$\displaystyle{{}##}$\hfil
      \crcr#1\crcr}}\,}
\fi
}

\AtBeginDocument{%
  \@ifpackageloaded{endfloat}%
   {\renewcommand\efloat@iwrite[1]{\immediate\expandafter\protected@write\csname efloat@post#1\endcsname{}}}{\newif\ifefloat@tables}%
}%

\def\BreakURLText#1{\@tfor\brk@tempa:=#1\do{\brk@tempa\hskip0pt}}
\let\lt=<
\let\gt=>
\def\processVert{\ifmmode|\else\textbar\fi}

\@ifundefined{subparagraph}{
\def\subparagraph{\@startsection{paragraph}{5}{2\parindent}{0ex plus 0.1ex minus 0.1ex}%
{0ex}{\normalfont\small\itshape}}%
}{}

\newcommand\role[1]{\unskip}
\newcommand\aucollab[1]{\unskip}
  
\@ifundefined{tsGraphicsScaleX}{\gdef\tsGraphicsScaleX{1}}{}
\@ifundefined{tsGraphicsScaleY}{\gdef\tsGraphicsScaleY{.9}}{}
\def\checkGraphicsWidth{\ifdim\Gin@nat@width>\linewidth
	\tsGraphicsScaleX\linewidth\else\Gin@nat@width\fi}

\def\checkGraphicsHeight{\ifdim\Gin@nat@height>.9\textheight
	\tsGraphicsScaleY\textheight\else\Gin@nat@height\fi}

\def\fixFloatSize#1{}
\let\ts@includegraphics\includegraphics

\def\inlinegraphic[#1]#2{{\edef\@tempa{#1}\edef\baseline@shift{\ifx\@tempa\@empty0\else#1\fi}\edef\tempZ{\the\numexpr(\numexpr(\baseline@shift*\f@size/100))}\protect\raisebox{\tempZ pt}{\ts@includegraphics{#2}}}}

\AtBeginDocument{\def\includegraphics{\@ifnextchar[{\ts@includegraphics}{\ts@includegraphics[width=\checkGraphicsWidth,height=\checkGraphicsHeight,keepaspectratio]}}}

\DeclareMathAlphabet{\mathpzc}{OT1}{pzc}{m}{it}

\def\URL#1#2{\@ifundefined{href}{#2}{\href{#1}{#2}}}

\def\UrlOrds{\do\*\do\-\do\~\do\'\do\"\do\-}%
\g@addto@macro{\UrlBreaks}{\UrlOrds}

\edef\fntEncoding{\f@encoding}

\makeatother

\newif\ifmultipleabstract\multipleabstractfalse%
\makeatletter

\def\wileyIndent{1pt}
\usepackage[paperheight=11in,paperwidth=8.5in,margin=2cm,headsep=.5cm,top=2.5cm,headheight=1cm]{geometry}

\renewenvironment{abstract}
{\vspace*{-1pc}\trivlist\item[]\leftskip\wileyIndent\hrulefill\par\vskip4pt\noindent\textbf{\abstractname}\mbox{\null}\\}{\par\noindent\hrulefill\endtrivlist}

\usepackage[]{footmisc}

\def\author#1{\gdef\@author{\hskip-\dimexpr(\tabcolsep)\hskip\wileyIndent\parbox{\dimexpr\textwidth-\wileyIndent}{\centering\bfseries#1}}}

\def\title#1{\linespread{1}\gdef\@title{\centering\bfseries\ifx\@articleType\@empty\else\@articleType\\\fi#1}}

\let\@articleType\@empty \def\articletype#1{\gdef\@articleType{{\normalfont\itshape#1}}}

 \def\audegree#1{}

\date{}

\makeatother

\usepackage[T1]{fontenc}
\makeatother
\usepackage[numbers,sort&compress]{natbib}
\setcitestyle{authoryear,open={(},close={)}}

\def\thanksspace{{\phantom{\textsuperscript{\thefootnote}}}}
\begin{document}

\title{Latent class growth analysis for ordinal response data in the Distress Assessment and Response Tool: an evaluation of state-of-the-art implementations}
\author{Jianhui~Gao\textsuperscript{1}\space,  Aliza Panjwani\textsuperscript{2}, Madeline Li\textsuperscript{2}, Osvaldo Espin-Garcia\textsuperscript{1,3}\thanks{E-mail:osvaldo.espingarcia@utoronto.ca}{\thanksspace}~\\[-3pt]\normalsize\normalfont  \itshape ~\\
\textsuperscript{1}{Division of Biostatistics, Dalla Lana School of Public Health\unskip, University of Toronto\unskip, Toronto\unskip, ON\unskip, Canada}~\\
\textsuperscript{2}{Department of Supportive Care, Princess Margaret Cancer Centre, University Health Network, Toronto\unskip, ON\unskip, Canada}~\\
\textsuperscript{3}{Department of Biostatistics, Princess Margaret Cancer Centre, University Health Network, Toronto\unskip, ON\unskip, Canada}}
\maketitle
\newpage

\noindent \textbf{Latent class growth analysis for ordinal response data in the Distress Assessment and Response Tool: an evaluation of state-of-the-art implementations}
\begin{abstract}
\def\keywordstitle{}
Latent class growth analysis is a popular approach to identify underlying subpopulations. Several implementations, such as LCGA (Mplus), Proc Traj (SAS) and lcmm (R) are specially designed for this purpose. Motivated by data collection of psychological instruments over time in a large North American cancer centre, we compare these implementations using various simulated Edmonton Symptom Assessment System revised (ESAS-r) scores, an ordinal outcome from 0 to 10, as well as the real data consisting of more than 20,000 patients. 
We found that Mplus and lcmm lead to high correct classification rate, but Proc Traj over estimated the number of classes and failed to converge. While Mplus is computationally faster than lcmm, it does not allow more than 10 levels. We therefore suggest first analyzing data on the ordinal scale using lcmm. If computational time becomes an issue, then one can group the scores into categories and implement them in Mplus. 
\end{abstract}
keywords: latent class growth analysis, ESAS-r, ordinal outcome

\newpage
\setlength\parindent{0pt}
\section*{Introduction}
The longitudinal study design has been involved in many areas of clinical psychology to observe and understand psychological behaviours of individuals over time. This study design is particularly useful for evaluating the relationship between risk factors and the development of disease, as well as treatment effects over different lengths of time \citep{caruana_longitudinal_2015}. In the past decade, longitudinal studies were often conducted in large populations across different clinic sites \citep[e.g.,][]{janelsins_cognitive_2017,livingstone_prospective_2015}. A latent class or growth mixture modeling approach  seems to be the most appropriate method for fully capturing information about interindividual differences in intraindividual change taking into account unobserved heterogeneity \citep{jung_introduction_2008}. Traditionally, such analysis is implemented in Mplus \citep{muthen_finite_1999} or SAS Proc Traj \citep{jones_sas_2001}. Recently, the package lcmm has enabled researchers to implement latent class mixture models freely in open-source software R \citep{proust-lima_estimation_2017}. \\

While recent studies \citep[e.g.,][]{nguena_nguefack_trajectory_2020} have described the three implementations in some detail along with their availability for data analysis, they have not been systematically evaluated in the literature. To the best of our knowledge, the tutorial by \cite{wardenaar_latent_2020} was the first paper explicitly evaluating lcmm and Mplus. Our study is different than these two studies in the following ways. \\

First, we have included SAS proc traj in our comparison as it is still a popular tool used to model group-based trajectory in recent-year clinical studies \citep{rubeis_group-based_2021}. \\

Second, motivated by depression and anxiety survey collected routinely through the Distress Assessment and Response Tool (DART) from the largest cancer center in Canada, we focus on a specific type of longitudinal outcome that is commonly used in quantitative symptom assessment: ordinal responses originating from numeric ratings \citep{johnson_novel_2015,hui_edmonton_2017}. Taking Edmonton Symptom Assessment System Revised (ESAS-r) as an example, ESAS-r scores of 0-3, 4-6 and 7-10 are generally considered as none/mild, moderate and severe in clinical practice \citep{seow_high_2012}. This original scale is not continuous nor normally distributed. In fact, many studies find a larger portion in the lowest category, which signifies an absence of the behavior and thus the distributions tend to be quite skewed \citep{feldman_new_2009}. Although Mplus (7.4) can analyze ordinal data using censored normal as well as cumulative probit models. it can only apply censoring to one side (i.e., censored-below or censored-above) and limits to a maximum of 10 levels for ordinal data (11 levels in ESAS-r). To work around this, we propose to first group the ordinal outcomes and before analyzing in Mplus. Both SAS proc traj and lcmm package can be used to analyze data on original scale. SAS proc traj provides three options: zero-inflated Poisson (ZIP), censored normal (CNORM) and logistic model, with CNORM preferred for psychometric scale data \citep{nagin_trajectories_1999}. Lcmm function from lcmm package uses a similarly cumulative probit model as Mplus, but it does not have restriction on number of ordinal levels. \\

Third, motivated by the large sample size (n=23,987) and number of surveys collected (N=122,901), we simulated 60,000 surveys across 12 time-points (compared to 500 surveys at 5 time-point in \cite{wardenaar_latent_2020}). This will bring the computational challenge for lcmm as we will see later. Furthermore, we evaluated each tool's performance regarding ability to detect the true number of latent class, posterior correct classification rate, and computational time, averaged across 1,000 independently simulated data-set (compared to only one data-set in \cite{wardenaar_latent_2020}). Because we are simulating 1,000 replicates (as opposed to a single synthetic dataset), we are able to better assess uncertainty associated with identifying the number of classes across implementations. This extensive simulation will allow us to examine the performance and convergence issue by default configuration. Although sometimes the non-convergence issue (in all 3 implementations) can be solved by manually searching and then setting a well-chosen start-point for model estimation, an implementation that often does not converge under default start-point will greatly limit clinician's enthusiasm to use. \\

Finally, to reveal how the challenges can arise in real data analysis, we study depression and anxiety scores on ESAS-r surveys routinely collected in patients with cancer through the Distress Assessment and Response Tool (DART) from the Princess Margaret Cancer Center between March 2013 and March 2018. 

\section*{Methods}
\subsection*{Model Assumptions}
For each individual $i$, let $\boldsymbol{y}_i = (y_{i1}, . . . , y_{iT})$ represent longitudinal trajectory over T periods and $\boldsymbol{z_i}={z_{i1},z_{i2},...,z_{ip}}$ be a set of baseline covariates. The key assumption in all models is conditional independence: for subject $i$, $\boldsymbol{z_i}$ and the data trajectory $\boldsymbol{y}_i$, are independent given the unobserved latent  group $C_i$. The likelihood of observing $\boldsymbol{y}_i$ given group $C_i=k$ is 
\begin{equation*}
\begin{split}
        f(\boldsymbol{y}_i\mid\boldsymbol{z_i})&=\sum_{i=1}^K Pr(C_i=k\mid\boldsymbol{z_i})Pr(\boldsymbol{y}_i\mid C_i=k)\\
        &=\sum_{i=1}^K\frac{\exp\left(\theta_k+\lambda_k^{T}z_i\right)}{\sum_{l=1}^k\exp\left(\theta_l+\lambda_l^{T}z_i\right)}Pr(\boldsymbol{y}_i\mid C_i=k),
\end{split}
\end{equation*}
where the group membership is determined by generalized logit function on covariates $z_i$. 
\subsection*{Likelihood}
Proc Traj CNORM models the conditional distribution as
\begin{equation*}
\begin{split}
    Pr(\boldsymbol{y}_i|C_i=k)=&\prod_{y_{ij}=Min}\Phi\Big(\frac{Min-\mu_{ijk}}{\sigma}\Big)\prod_{Min<y_{ij}<Max}\frac{1}{\sigma}\Phi\Big(\frac{y_{ij}-\mu_{ijk}}{\sigma}\Big)\times\\
    &\prod_{y_{ij}=Max}\Big(1-\Phi(\frac{Max-\mu_{ijk}}{\sigma})\Big),
\end{split}
\end{equation*}
where $\Phi$ is the standard normal distribution and $\sigma^2$ is the variance of $\boldsymbol{y_i}$. In our case values are bounded between $Min = 0$ and $Max = 10$. \\

To model ordinal outcome with $M$ levels, both lcmm package and Mplus use generalized probit model with cumulative probability
\begin{equation*}
    Pr(y_{ij} \le l|C_i =k)=\Phi(\eta_l - \mu_{ijk}) \text{ }l = 1,....M.
\end{equation*}
In order to ensure well-defined probabilities, we require that $\eta_l > \eta_{l-1}$, $\forall l$, and it is understood that $\eta_M=\infty$  such that $\Phi(\infty) = 1$ as well as $\eta_0 = -\infty$ such that $\Phi(-\infty)=0$. This model can be motivated by a latent continous variable $y_{ij}^*|k=\mu_{ijk}+\epsilon_{ik}$ following: $y_{ij}|k = l \iff \eta_{l-1}\le y_{ij}^*|k <\eta_{l}$ \citep{boes_ordered_2006}. Likelihood can therefore be formulated as
\begin{equation*}
Pr(\boldsymbol{y}_i|C_i=k)= \prod_{j=1}^T\prod_{l=1}^{M}(\Phi(\eta_{l}-\mu_{ijk})-\Phi(\eta_{l-1} - \mu_{ijk}))^{\boldsymbol{1}_{\{y_{ij}=l\}}},
\end{equation*}
where $\eta_1 < \eta_2 <...<\eta_{M-1}$ are cut points defined above.\\

Maximum likelihood estimation (MLE) is then performed by each software with different procedures. SAS uses the general quasi-Newton procedure with default initial value calculated on group intercepts spaced. Mplus, on the other hand, uses a combination of  Expectation-Maximization and quasi-Newton algorithm with twenty sets of random starting points. We use recommended automatic specification from one-class model estimates as initial values, but random draws from asymptotic distribution of MLE of one-class are also possible. Lcmm solves MLE by Iterative Marquardt algorithm. The convergence criteria and rate are to be discussed in later sections. A summary of main features can be found in Table \ref{tab:summary}.

\subsection*{Simulation Study Design}
We assign three time-stable covariates values to 5,000 patients and generate ESAS-r scores from these covariates over 12 equally spaced time periods. Patients with different covariates were assigned to exhibit three distinct trajectory patterns: 3,000 patients have constant ESAS-r low scores, 1,000 patients have increasing scores, and the remaining 1000 have decreasing ESAS-r over time. Let $\epsilon_t \sim N(0,1)$ be a random noise and $Y_{it} \sim Exp(\lambda_t) \text{ with }\lambda_t = z_1+0.5*z_2*t-0.05*z_3*t+\epsilon$. We round $y_{it}$ to nearest integer and bound them between 0 and 10 to mimic an ESAS-r value. Constant low, increasing and decreasing groups are generated by setting $z_i = (2,0,0), (1,0,1)$, or $(0,1,0)$, respectively. A graphical illustration of a particular set of simulated trajectories is shown in Figure \ref{fig:siml}.\\

Oncology patients can experience events such as financial hardship, improved health conditions or cancer recurrence, which sometimes cannot be observed yet can have significant impacts on their psychosocial state. In a second set of stimulation studies, we assume everyone experiences an event with  impact on reducing ESAS-r scores, for example, treatment. Let $T_{it}$ be a event time for $i^{th}$ individual, and $Y_{it}\sim Exp(\lambda_t)$ with $\lambda_t=z_1+0.01*z_2*t-0.02*z_3*t+\boldsymbol{1}_{\{T_{it}>t\}}*\beta_{tr}+\epsilon_t$. We simulate $T_{it} \sim NegBin(1,0.5) + 1$ to mimic life-like scenarios in which patients are more likely to receive treatment closer to diagnosis/baseline. Constantly low, increasing and decreasing groups are generated by $z_i = (0.2,0,0),(0.3,0,1),(0.15,1,0)$ and $\beta_{tr}$ = 0.2 respectively. Figure \ref{fig:sim2} shows changes in average trajectory when adding this treatment effect. The purpose of this second set of simulations is to evaluate the effect of an unmeasured variable in the implementations' performance\\

We first analyze the data-set with lcmm and Proc Traj on the original scale. Then data are converted to three ordered categories: none/mild(0-3), moderate(4-6), severe(7-10) and analyzed in Mplus and R lcmm. To determine the best number of latent classes, we have used the Bayesian Information Criterion (BIC) defined as $BIC = -2\log L + p \log(n)$. Extensive simulation studies have shown that BIC performs better than Lo-Mendell-Rubin likelihood ratio test and other statistical information criteria (IC), such as Akaike’s Information Criterion (AIC) and sample size adjusted BIC \citep{nylund_deciding_2007}. Although \cite{nylund_deciding_2007} show in their simulations that bootstrap likelihood ratio test (BLRT) has advantages over BIC, BLRT is not considered in this paper because lcmm and SAS do not provide this option and BLRT requires considerable additional computational time. We use logged Bayes factor ($2\Delta$BIC) between classes as evidence against null model \citep{jones_sas_2001}. We simulated 1,000 replicates in total. Computations were performed in Mplus (7.4) and SAS (9.4) using a local laptop (i7-3770K CPU @ 3.50GHz, 4 Core(s), and 8GB RAM). Convserly, R lcmm(1.9.2) was executed on the Niagara supercomputer (20 nodes used, each node consists of 40 Intel "Skylake" @ 2.4 GHz) at the SciNet HPC Consortium \citep{loken_scinet_2010}.

\subsection*{Distress Assessment and Response Tool}
The Distress Assessment and Response Tool (DART) is an innovative program developed by the Department of Supportive Care at Princess Margaret Cancer Center (PM) \citep{li_easier_2016}. Every patient at PM has the opportunity to complete the self-assessment on touch-screen tables in waiting areas before their appointment. In this study, we focus on the standardized Edmonton Symptom Assessment System Revised (ESAS-r) depression and anxiety scores from DART. This data-set includes surveys completed between 2013 and 2018 across eleven oncology sites: bone marrow transplantation/hematology, breast, gastrointestinal, genitourinary, gynecology, head and neck, leukemia, lung, lymphoma, multiple myeloma, and sarcoma. We exclude surveys missing any of the following clinical measures: depression score, anxiety score, ECOG performance status or stage of cancer as well as missing baseline characteristics such as age, sex and household income information. To balance our dataset, we limit longitudinal repeated measures to 12 for each patient. After these exclusion criteria were applied, 122,901 surveys completed by 23,987 patients remained for further analyses. We fit them using  Proc Traj and lcmm on the original scale and using Mplus and lcmm on grouped categories. 

\section*{Results}
\subsection*{Simulations}
Number of classes identified by each model according to $\Delta BIC$ is highlighted in red in Table \ref{tab:BIC}. Without considering the unmeasured treatment effect, almost all algorithms can correctly identify the number of latent classes, except for Proc Traj. Once treatment effect is added, however, only R lcmm with the original scale and Mplus can identified the number of groups properly (Table 2, bottom). Table \ref{tab:pprob} shows the average posterior classification rate if the true number of classes is known. Without treatment effects, all implementations perform well on classifying patients to the constantly low group ($>92\%$). This number can decrease to less than 70\% for the other two groups if covariates are not included. Correct classification rate is generally lower with the presence of treatment effect, yet Mplus and lcmm can still maintain $90\%$ or higher for all groups. Mplus and lcmm have consistently high correct classification rates while exhibiting ability to detect the true number of latent classes across various simulation scenarios. Table \ref{tab:computational time} shows average computational time in simulation studies. Lcmm is considerably more computationally intensive than SAS Proc Traj and Mplus, particularly if models are using the original scale. SAS Proc Traj and Mplus usually take less than few minutes to run, whereas it can take up to a few hours for lcmm. 

\subsection*{DART}
Table \ref{tab:convergence} shows BIC of fitting DART data by Proc Traj on different number of classes, AIC as well as log-likelihood are also included for comparison. SAS failed to converge at 5 groups for anxiety and 6 groups for depression, and therefore were excluded for model selection. Based on the remaining BICs, 4 distinct groups are identified for anxiety and additional two groups for depression (Figure \ref{fig:sas_traj}). Mean score for patients in all but group 5 are roughly the same over time. Patients in group 5 exhibit a slightly decreasing trend from 1 to 0. \\

Mplus, on the other hand, has the lowest BIC value at 5 groups for anxiety and 4 groups for depression (Table \ref{tab:convergence}). The result is harder to interpret than before because of cumulative probit structure on categorical responses.
Figures \ref{fig:anxiety_prob}-\ref{fig:depression_prob} show estimated probability of each group stratified by anxiety/depression categories. For example, 59.9\% of the patients are assigned to group 3 (Figures \ref{fig:anxiety_prob}) according to their anxiety trajectories. This group has nearly 100\% probability of having none/mild anxiety at any given time. Group 3, therefore, is equivalent to the constantly low group. Similarly group 2 and group 5 represent patients that always exhibit high and moderate anxiety respectively. Patients in group 1 on average improved their anxiety as they have increasing probability of experiencing none/mild over time. On the contrary, patients in group 4 have worsened anxiety over time. The group assignment is very similar for depression (Figure \ref{fig:depression_prob}) except that the last group (group 5) is not not present. 

\section*{Conclusion}

R lcmm, SAS Proc Traj and Mplus can all run latent growth analysis on psychological assessments that scale between 0 and 10. R lcmm assumes data are ordinal and thus uses generalized probit function, whereas SAS Proc Traj makes the assumption that data are continuous and normal (censored). Both models can be implemented in Mplus, however, ordinal levels are restricted to a maximum of 10 and censoring can only be done at one side. A usual way to get around these restrictions for Mplus is to group several scores together, but collapsing ordinal data into fewer categories does result in loss of information.\\

In our simulation study, we have demonstrated that when all covariates can be identified, each software performs similarly well. As we move towards more complex simulations (when unmeasured time-varying latent events are considered), Proc Traj over estimates the number of classes in many simulation settings and has the lowest posterior classification rate on average. Both Mplus and lcmm have high posterior classification rates and correctly identify the number of latent classes in at least one scenario. The computational time of lcmm, usually a few hours with logit link, makes it less appealing in large scale data applications. Lcmm becomes less feasible to run as the number of classes increases, especially when covariate selection is also needed. We hypothesize that this extra computation time maybe due in part to the strict convergence criteria imposed by lcmm. Convergence in Mplus and Proc Traj are based on log-likelihood, whereas all three convergence criteria on parameter, log-likelihood and gradient must be simultaneously satisfied in lcmm to ensure good convergence \citep{proust-lima_estimation_2017}. However it is not evident from our simulation that lcmm has a higher posterior classification rate than Mplus. \\

With the application to one of the largest psychological assessment datasets in North America, Proc Traj is a natural choice because it can run directly on the original scale within few minutes. But Proc Traj fails to converge past a certain number of groups and this convergence issue has been repeatedly reported repeatedly in the literature \citep{wiesner_arrest_2007,kj_substance_2016,genolini_kml_2010}. \cite{blaze_enumerating_2014} has observed a non-convergence pattern when the class mixing proportions were extremely unequal and the sample size was large. Lcmm is not attempted here for the reasons outlined above and instead Mplus is used on  grouped categories. Mplus finds different trajectory patterns although more than 60\% of patients were assigned to same group as Proc Traj. Most disagreement happens between none/mild and medium group, which could occur due to the convergence issue of Proc Traj. We have limited our studies to the most basic features of latent class growth analysis and we acknowledge that lcmm and Mplus can be further explored by incorporating random structure within class to fit longitudinal data. However, this additional exploration may be even more computationally expensive and prohibitive with our current sample size. For ordinal outcome data like ESAS-r, we recommend first to use lcmm directly on the ordinal scale. If the computational time becomes an issue, then one can use Mplus with probit model if the ordinal outcome has $\le 10$ levels or grouping the levels into fewer categories. 

\section*{Data Availability:} Data Availability: All data used in the simulations can be reproduced from [removed for peer review].The clinical data that support the findings of this study are available on request from the corresponding author, [removed for peer review]]. The data are not publicly available due to containing information that could compromise the privacy of research participants.

\medskip
\bibliography{reference}

@article{nguena_nguefack_trajectory_2020,
	title = {Trajectory {Modelling} {Techniques} {Useful} to {Epidemiological} {Research}: {A} {Comparative} {Narrative} {Review} of {Approaches}},
	volume = {12},
	issn = {1179-1349},
	shorttitle = {Trajectory {Modelling} {Techniques} {Useful} to {Epidemiological} {Research}},
	url = {https://www.ncbi.nlm.nih.gov/pmc/articles/PMC7608582/},
	doi = {10.2147/CLEP.S265287},
	urldate = {2021-07-16},
	journal = {Clinical Epidemiology},
	author = {Nguena Nguefack, Hermine Lore and Pagé, M Gabrielle and Katz, Joel and Choinière, Manon and Vanasse, Alain and Dorais, Marc and Samb, Oumar Mallé and Lacasse, Anaïs},
	month = oct,
	year = {2020},
	pmid = {33154677},
	pmcid = {PMC7608582},
	pages = {1205--1222},
}

@techreport{wardenaar_latent_2020,
	title = {Latent {Class} {Growth} {Analysis} and {Growth} {Mixture} {Modeling} using {R}: {A} tutorial for two {R}-packages and a comparison with {Mplus}.},
	shorttitle = {Latent {Class} {Growth} {Analysis} and {Growth} {Mixture} {Modeling} using {R}},
	url = {https://psyarxiv.com/m58wx/},
	urldate = {2021-07-16},
	institution = {PsyArXiv},
	author = {Wardenaar, Klaas},
	month = apr,
	year = {2020},
	doi = {10.31234/osf.io/m58wx},
	note = {type: article},
}

@article{rubeis_group-based_2021,
	title = {Group-based trajectory modeling of body mass index and body size over the life course: {A} scoping review},
	volume = {7},
	issn = {2055-2238},
	shorttitle = {Group-based trajectory modeling of body mass index and body size over the life course},
	url = {https://onlinelibrary.wiley.com/doi/abs/10.1002/osp4.456},
	doi = {10.1002/osp4.456},
	language = {en},
	number = {1},
	urldate = {2021-07-16},
	journal = {Obesity Science \& Practice},
	author = {Rubeis, Vanessa De and Andreacchi, Alessandra T. and Sharpe, Isobel and Griffith, Lauren E. and Keown-Stoneman, Charles D. G. and Anderson, Laura N.},
	year = {2021},
	note = {\_eprint: https://onlinelibrary.wiley.com/doi/pdf/10.1002/osp4.456},
	pages = {100--128},
}

@article{livingstone_prospective_2015,
	title = {Prospective evaluation of follow-up in melanoma patients in {Germany} – {Results} of a multicentre and longitudinal study},
	volume = {51},
	issn = {0959-8049},
	url = {http://www.sciencedirect.com/science/article/pii/S0959804915000155},
	doi = {10.1016/j.ejca.2015.01.007},
	language = {en},
	number = {5},
	urldate = {2020-12-17},
	journal = {European Journal of Cancer},
	author = {Livingstone, E. and Krajewski, C. and Eigentler, T. K. and Windemuth-Kieselbach, C. and Benson, S. and Elsenbruch, S. and Hauschild, A. and Rompel, R. and Meiss, F. and Mauerer, A. and Kähler, K. C. and Dippel, E. and Möllenhoff, K. and Kilian, K. and Mohr, P. and Utikal, J. and Schadendorf, D.},
	month = mar,
	year = {2015},
	pages = {653--667},
}

@article{janelsins_cognitive_2017,
	title = {Cognitive {Complaints} in {Survivors} of {Breast} {Cancer} {After} {Chemotherapy} {Compared} {With} {Age}-{Matched} {Controls}: {An} {Analysis} {From} a {Nationwide}, {Multicenter}, {Prospective} {Longitudinal} {Study}},
	volume = {35},
	issn = {0732-183X},
	shorttitle = {Cognitive {Complaints} in {Survivors} of {Breast} {Cancer} {After} {Chemotherapy} {Compared} {With} {Age}-{Matched} {Controls}},
	url = {https://www.ncbi.nlm.nih.gov/pmc/articles/PMC5455314/},
	doi = {10.1200/JCO.2016.68.5826},
	number = {5},
	urldate = {2020-12-17},
	journal = {Journal of Clinical Oncology},
	author = {Janelsins, Michelle C. and Heckler, Charles E. and Peppone, Luke J. and Kamen, Charles and Mustian, Karen M. and Mohile, Supriya G. and Magnuson, Allison and Kleckner, Ian R. and Guido, Joseph J. and Young, Kelley L. and Conlin, Alison K. and Weiselberg, Lora R. and Mitchell, Jerry W. and Ambrosone, Christine A. and Ahles, Tim A. and Morrow, Gary R.},
	month = feb,
	year = {2017},
	pmid = {28029304},
	pmcid = {PMC5455314},
	pages = {506--514},
}

@article{caruana_longitudinal_2015,
	title = {Longitudinal studies},
	volume = {7},
	issn = {2072-1439},
	url = {https://www.ncbi.nlm.nih.gov/pmc/articles/PMC4669300/},
	doi = {10.3978/j.issn.2072-1439.2015.10.63},
	number = {11},
	urldate = {2020-12-17},
	journal = {Journal of Thoracic Disease},
	author = {Caruana, Edward Joseph and Roman, Marius and Hernández-Sánchez, Jules and Solli, Piergiorgio},
	month = nov,
	year = {2015},
	pmid = {26716051},
	pmcid = {PMC4669300},
	pages = {E537--E540},
}

@article{genolini_kml_2010,
	title = {{KmL}: k-means for longitudinal data},
	volume = {25},
	issn = {0943-4062, 1613-9658},
	shorttitle = {{KmL}},
	url = {http://link.springer.com/10.1007/s00180-009-0178-4},
	doi = {10.1007/s00180-009-0178-4},
	language = {en},
	number = {2},
	urldate = {2020-10-22},
	journal = {Computational Statistics},
	author = {Genolini, Christophe and Falissard, Bruno},
	month = jun,
	year = {2010},
	pages = {317--328},
}

@article{li_easier_2016,
	title = {Easier {Said} {Than} {Done}: {Keys} to {Successful} {Implementation} of the {Distress} {Assessment} and {Response} {Tool} ({DART}) {Program}},
	volume = {12},
	issn = {1554-7477},
	shorttitle = {Easier {Said} {Than} {Done}},
	url = {https://ascopubs.org/doi/full/10.1200/JOP.2015.010066},
	doi = {10.1200/JOP.2015.010066},
	number = {5},
	urldate = {2020-10-31},
	journal = {Journal of Oncology Practice},
	author = {Li, Madeline and Macedo, Alyssa and Crawford, Sean and Bagha, Sabira and Leung, Yvonne W. and Zimmermann, Camilla and Fitzgerald, Barbara and Wyatt, Martha and Stuart-McEwan, Terri and Rodin, Gary},
	month = apr,
	year = {2016},
	note = {Publisher: American Society of Clinical Oncology},
	pages = {e513--e526},
}

@phdthesis{blaze_enumerating_2014,
	title = {Enumerating the {Correct} {Number} of {Classes} in a {Semiparametric} {Group}-based {Trajectory} {Model}},
	language = {en},
	author = {Blaze, Thomas James},
	year = {2014},
}

@article{nylund_deciding_2007,
	title = {Deciding on the {Number} of {Classes} in {Latent} {Class} {Analysis} and {Growth} {Mixture} {Modeling}: {A} {Monte} {Carlo} {Simulation} {Study}},
	volume = {14},
	issn = {1070-5511, 1532-8007},
	shorttitle = {Deciding on the {Number} of {Classes} in {Latent} {Class} {Analysis} and {Growth} {Mixture} {Modeling}},
	url = {https://www.tandfonline.com/doi/full/10.1080/10705510701575396},
	doi = {10.1080/10705510701575396},
	language = {en},
	number = {4},
	urldate = {2020-10-22},
	journal = {Structural Equation Modeling: A Multidisciplinary Journal},
	author = {Nylund, Karen L. and Asparouhov, Tihomir and Muthén, Bengt O.},
	month = oct,
	year = {2007},
	pages = {535--569},
}

@article{kj_substance_2016,
	title = {Substance {Use} {Trajectories} {From} {Early} {Adolescence} {Through} the {Transition} to {College}.},
	volume = {77},
	issn = {1937-1888, 1938-4114},
	url = {http://europepmc.org/article/PMC/5088174},
	doi = {10.15288/jsad.2016.77.924},
	language = {English},
	number = {6},
	urldate = {2020-10-22},
	journal = {Journal of Studies on Alcohol and Drugs},
	author = {Kj, Derefinko and Rj, Charnigo and Jr, Peters and Zw, Adams and R, Milich and Dr, Lynam},
	month = nov,
	year = {2016},
	pmid = {27797694},
	pages = {924--935},
}

@article{wiesner_arrest_2007,
	title = {Arrest {Trajectories} {Across} a 17-{Year} {Span} for {Young} {Men}: {Relation} to {Dual} {Taxonomies} and {Self}-{Reported} {Offense} {Trajectories}},
	volume = {45},
	issn = {0011-1384},
	shorttitle = {Arrest {Trajectories} {Across} a 17-{Year} {Span} for {Young} {Men}},
	url = {https://www.ncbi.nlm.nih.gov/pmc/articles/PMC2600715/},
	doi = {10.1111/j.1745-9125.2007.00099.x},
	number = {4},
	urldate = {2020-10-22},
	journal = {Criminology; an interdisciplinary journal},
	author = {Wiesner, Margit and Capaldi, Deborah M. and Kim, Hyoun K.},
	month = nov,
	year = {2007},
	pmid = {19079783},
	pmcid = {PMC2600715},
	pages = {835--863},
}

@article{loken_scinet_2010,
	title = {{SciNet}: {Lessons} {Learned} from {Building} a {Power}-efficient {Top}-20 {System} and {Data} {Centre}},
	volume = {256},
	issn = {1742-6596},
	shorttitle = {{SciNet}},
	url = {https://doi.org/10.1088%2F1742-6596%2F256%2F1%2F012026},
	doi = {10.1088/1742-6596/256/1/012026},
	language = {en},
	urldate = {2020-09-09},
	journal = {Journal of Physics: Conference Series},
	author = {Loken, Chris and Gruner, Daniel and Groer, Leslie and Peltier, Richard and Bunn, Neil and Craig, Michael and Henriques, Teresa and Dempsey, Jillian and Yu, Ching-Hsing and Chen, Joseph and Dursi, L. Jonathan and Chong, Jason and Northrup, Scott and Pinto, Jaime and Knecht, Neil and Zon, Ramses Van},
	month = nov,
	year = {2010},
	note = {Publisher: IOP Publishing},
	pages = {012026},
}

@article{nagin_trajectories_1999,
	title = {Trajectories of boys' physical aggression, opposition, and hyperactivity on the path to physically violent and nonviolent juvenile delinquency},
	volume = {70},
	issn = {0009-3920},
	doi = {10.1111/1467-8624.00086},
	language = {eng},
	number = {5},
	journal = {Child Development},
	author = {Nagin, D. and Tremblay, R. E.},
	month = oct,
	year = {1999},
	pmid = {10546339},
	pages = {1181--1196},
}

@article{feldman_new_2009,
	title = {New {Approaches} to {Studying} {Problem} {Behaviors}: {A} {Comparison} of {Methods} for {Modeling} {Longitudinal}, {Categorical} {Adolescent} {Drinking} {Data}},
	volume = {45},
	issn = {0012-1649},
	shorttitle = {New {Approaches} to {Studying} {Problem} {Behaviors}},
	url = {https://www.ncbi.nlm.nih.gov/pmc/articles/PMC2791967/},
	doi = {10.1037/a0014851},
	number = {3},
	urldate = {2020-08-06},
	journal = {Developmental psychology},
	author = {Feldman, Betsy J. and Masyn, Katherine E. and Conger, Rand D.},
	month = may,
	year = {2009},
	pmid = {19413423},
	pmcid = {PMC2791967},
	pages = {652--676},
}

@article{seow_high_2012,
	title = {Do {High} {Symptom} {Scores} {Trigger} {Clinical} {Actions}? {An} {Audit} {After} {Implementing} {Electronic} {Symptom} {Screening}},
	volume = {8},
	issn = {1554-7477},
	shorttitle = {Do {High} {Symptom} {Scores} {Trigger} {Clinical} {Actions}?},
	url = {https://www.ncbi.nlm.nih.gov/pmc/articles/PMC3500488/},
	doi = {10.1200/JOP.2011.000525},
	number = {6},
	urldate = {2020-08-06},
	journal = {Journal of Oncology Practice},
	author = {Seow, Hsien and Sussman, Jonathan and Martelli-Reid, Lorraine and Pond, Greg and Bainbridge, Daryl},
	month = nov,
	year = {2012},
	pmid = {23598849},
	pmcid = {PMC3500488},
	pages = {e142--e148},
}

@article{proust-lima_estimation_2017,
	title = {Estimation of {Extended} {Mixed} {Models} {Using} {Latent} {Classes} and {Latent} {Processes}: {The} {R} {Package} lcmm},
	volume = {78},
	copyright = {Copyright (c) 2017 Cécile Proust-Lima, Viviane Philipps, Benoit Liquet},
	issn = {1548-7660},
	shorttitle = {Estimation of {Extended} {Mixed} {Models} {Using} {Latent} {Classes} and {Latent} {Processes}},
	url = {https://www.jstatsoft.org/index.php/jss/article/view/v078i02},
	doi = {10.18637/jss.v078.i02},
	language = {en},
	number = {1},
	urldate = {2020-08-06},
	journal = {Journal of Statistical Software},
	author = {Proust-Lima, Cécile and Philipps, Viviane and Liquet, Benoit},
	month = jun,
	year = {2017},
	note = {Number: 1},
	pages = {1--56},
}

@article{muthen_finite_1999,
	title = {Finite mixture modeling with mixture outcomes using the {EM} algorithm},
	volume = {55},
	issn = {0006-341X},
	doi = {10.1111/j.0006-341x.1999.00463.x},
	language = {eng},
	number = {2},
	journal = {Biometrics},
	author = {Muthén, B. and Shedden, K.},
	month = jun,
	year = {1999},
	pmid = {11318201},
	pages = {463--469},
}

@article{jung_introduction_2008,
	title = {An {Introduction} to {Latent} {Class} {Growth} {Analysis} and {Growth} {Mixture} {Modeling}},
	volume = {2},
	issn = {1751-9004, 1751-9004},
	url = {http://doi.wiley.com/10.1111/j.1751-9004.2007.00054.x},
	doi = {10.1111/j.1751-9004.2007.00054.x},
	language = {en},
	number = {1},
	urldate = {2020-08-06},
	journal = {Social and Personality Psychology Compass},
	author = {Jung, Tony and Wickrama, K. A. S.},
	month = jan,
	year = {2008},
	pages = {302--317},
}

@article{jones_sas_2001,
	title = {A {SAS} {Procedure} {Based} on {Mixture} {Models} for {Estimating} {Developmental} {Trajectories}},
	volume = {29},
	issn = {0049-1241, 1552-8294},
	url = {http://journals.sagepub.com/doi/10.1177/0049124101029003005},
	doi = {10.1177/0049124101029003005},
	language = {en},
	number = {3},
	urldate = {2020-01-18},
	journal = {Sociological Methods \& Research},
	author = {Jones, Bobby L. and Nagin, Daniel S. and Roeder, Kathryn},
	month = feb,
	year = {2001},
	pages = {374--393},
}

@article{boes_ordered_2006,
	title = {Ordered response models},
	volume = {90},
	issn = {1614-0176},
	url = {https://doi.org/10.1007/s10182-006-0228-y},
	doi = {10.1007/s10182-006-0228-y},
	language = {en},
	number = {1},
	urldate = {2021-01-15},
	journal = {Allgemeines Statistisches Archiv},
	author = {Boes, Stefan and Winkelmann, Rainer},
	month = mar,
	year = {2006},
	pages = {167--181},
}

@article{hui_edmonton_2017,
	title = {The {Edmonton} {Symptom} {Assessment} {System} 25 {Years} {Later}: {Past}, {Present} and {Future} {Developments}},
	volume = {53},
	issn = {0885-3924},
	shorttitle = {The {Edmonton} {Symptom} {Assessment} {System} 25 {Years} {Later}},
	url = {https://www.ncbi.nlm.nih.gov/pmc/articles/PMC5337174/},
	doi = {10.1016/j.jpainsymman.2016.10.370},
	number = {3},
	urldate = {2021-03-12},
	journal = {Journal of pain and symptom management},
	author = {Hui, David and Bruera, Eduardo},
	month = mar,
	year = {2017},
	pmid = {28042071},
	pmcid = {PMC5337174},
	pages = {630--643},
}

@article{johnson_novel_2015,
	title = {A {Novel} {Quantitative} {Pain} {Assessment} {Instrument} {That} {Provides} {Means} of {Comparing} {Patient}’s {Pain} {Magnitude} {With} a {Measurement} of {Their} {Pain} {Tolerance}},
	volume = {7},
	issn = {1918-3003},
	url = {https://www.ncbi.nlm.nih.gov/pmc/articles/PMC4554218/},
	doi = {10.14740/jocmr2277e},
	number = {10},
	urldate = {2021-03-12},
	journal = {Journal of Clinical Medicine Research},
	author = {Johnson, Lanny L. and Pittsley, Andrew and Becker, Ruth and Young, Allison De},
	month = oct,
	year = {2015},
	pmid = {26346200},
	pmcid = {PMC4554218},
	pages = {781--790},
}
\newpage
\section*{Tables and figures}
\begin{figure}[htb!]
    \centering
    \includegraphics[width=0.8\linewidth]{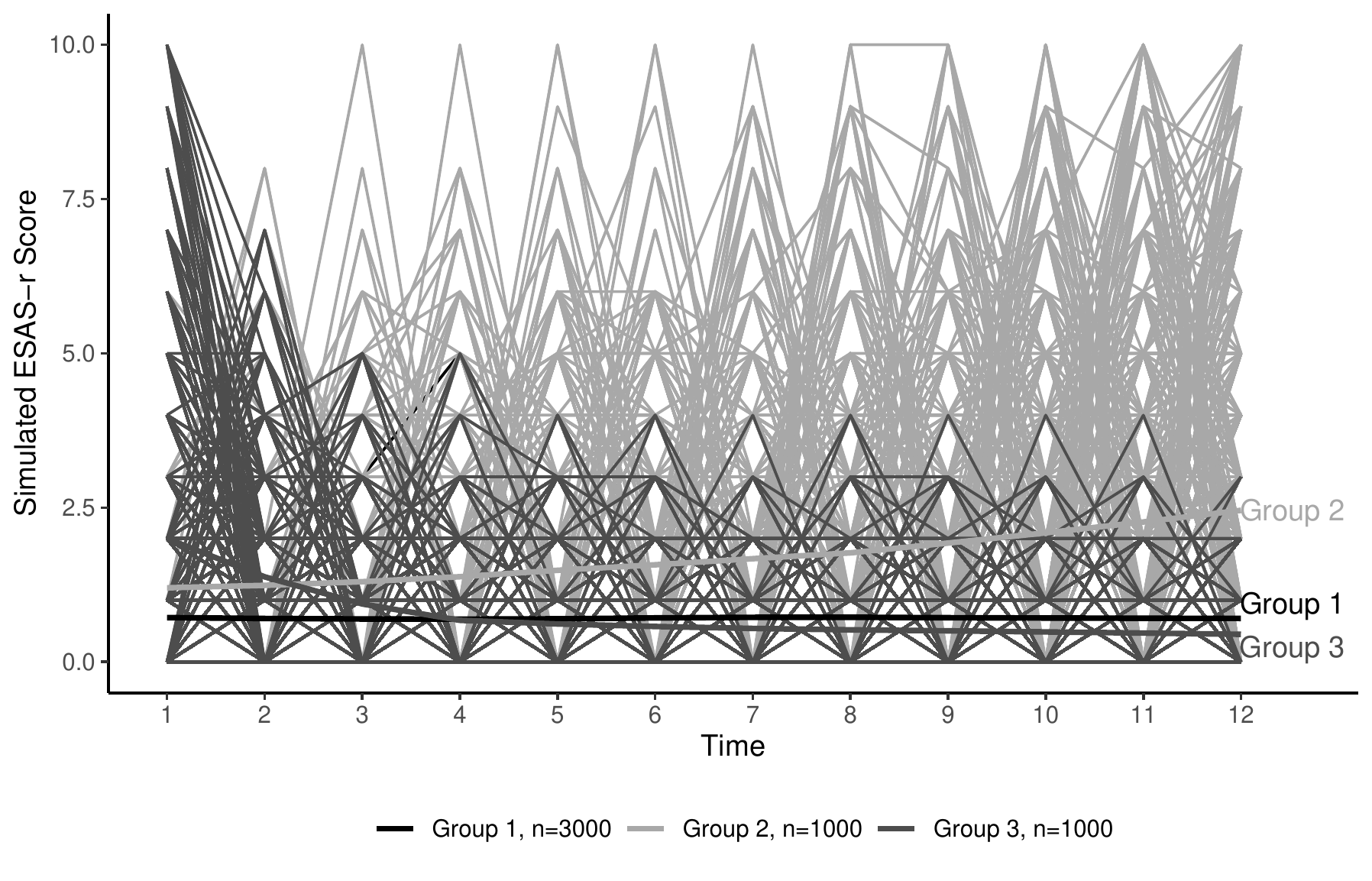}
    \caption{Simulated Trajectories of 5,000 patients' ESAS-r scores over 12 equally spaced time periods. Each line represents an individual's trajectory and the latent group they belong to. Lines with lables denote the average trajectory development within a group.}
    \label{fig:siml}
\end{figure}

\clearpage
\begin{figure}[htb!]
    \centering
    \includegraphics[width=\linewidth]{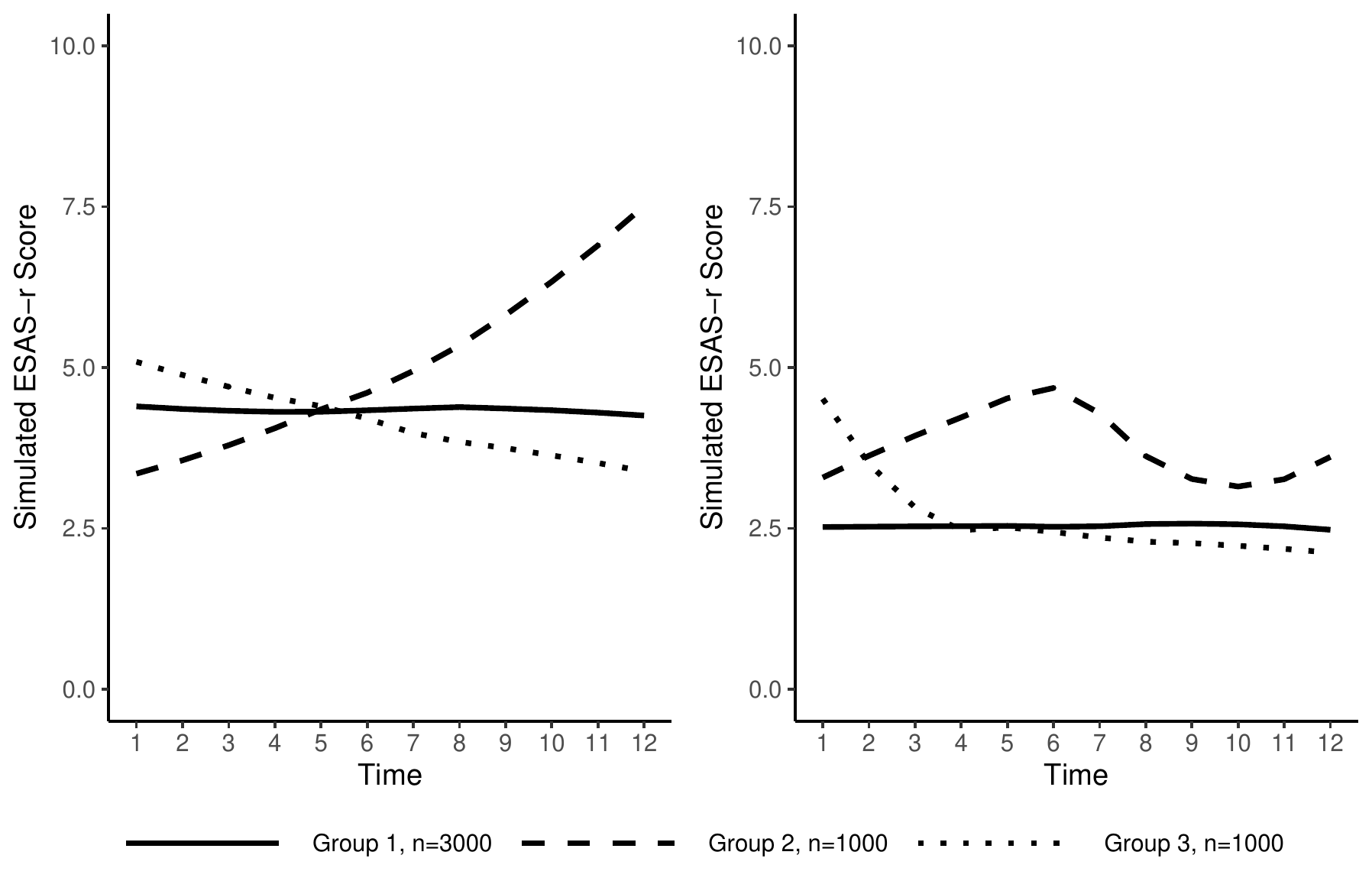}
    \caption{Average trajectories of ESAS-r scores over 12 equally spaced time periods without treatment effect (Left). Average trajectories on same patients with treatment interaction (Right). We assume everyone receives a treatment and reduces ESAS-r scores after treatment. ESAS-r scores are simulated on 5,000 patients independently. }
    \label{fig:sim2}
\end{figure}
\clearpage

\begin{figure}[htb!]
    \centering
    \includegraphics[width=\linewidth]{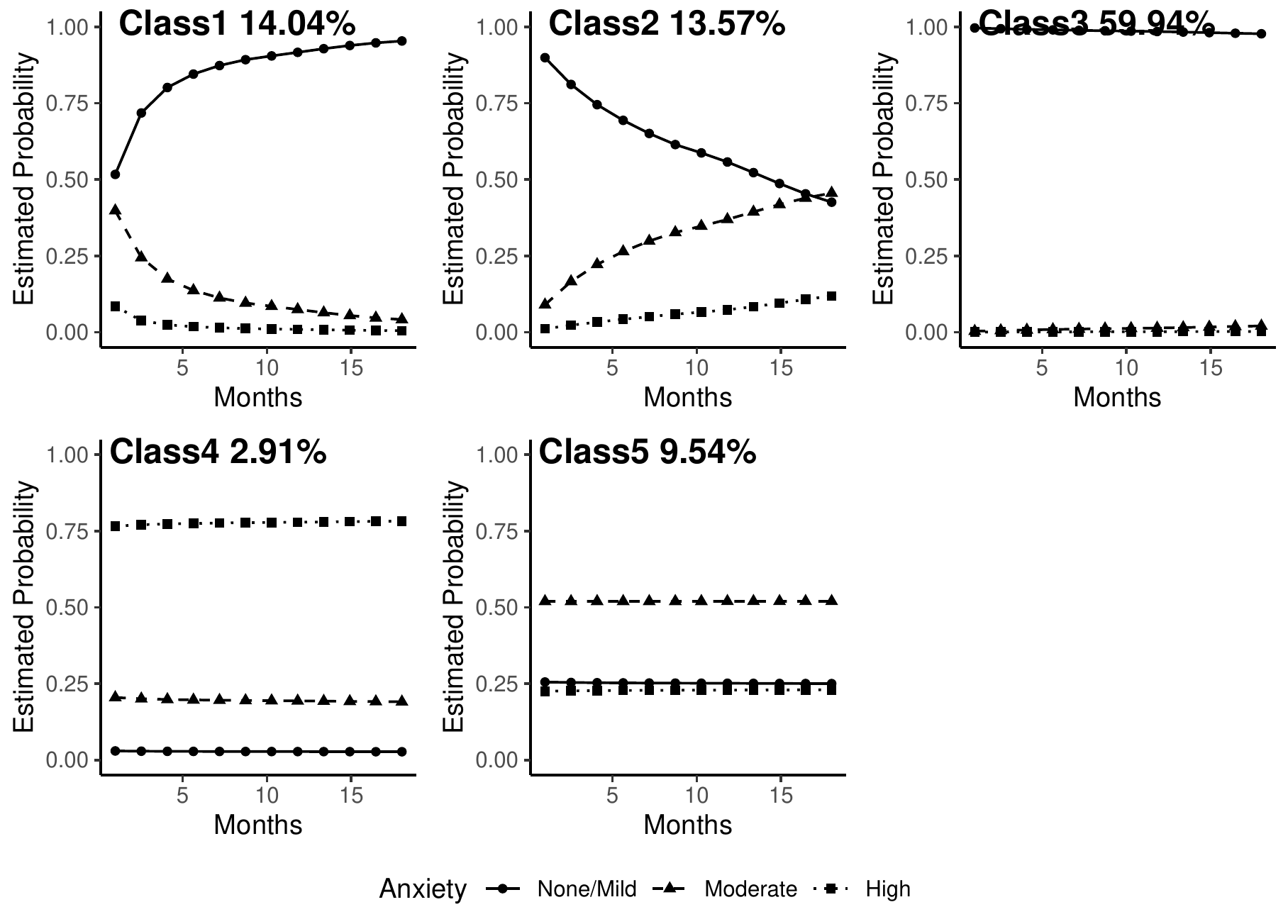}
    \caption{Estimated probabilities of each latent group found on DART ESAS-r Anxiety servery. Five distinct classes are found. Anxiety scores are grouped into None/Mild(0-3), Moderate(4-6) and High(7-10). Each line represents the probability for a patient in a given latent class falling into one of the anxiety categories over time.}
    \label{fig:anxiety_prob}
\end{figure}

\begin{figure}[htb!]
    \centering
    \includegraphics[width=\linewidth]{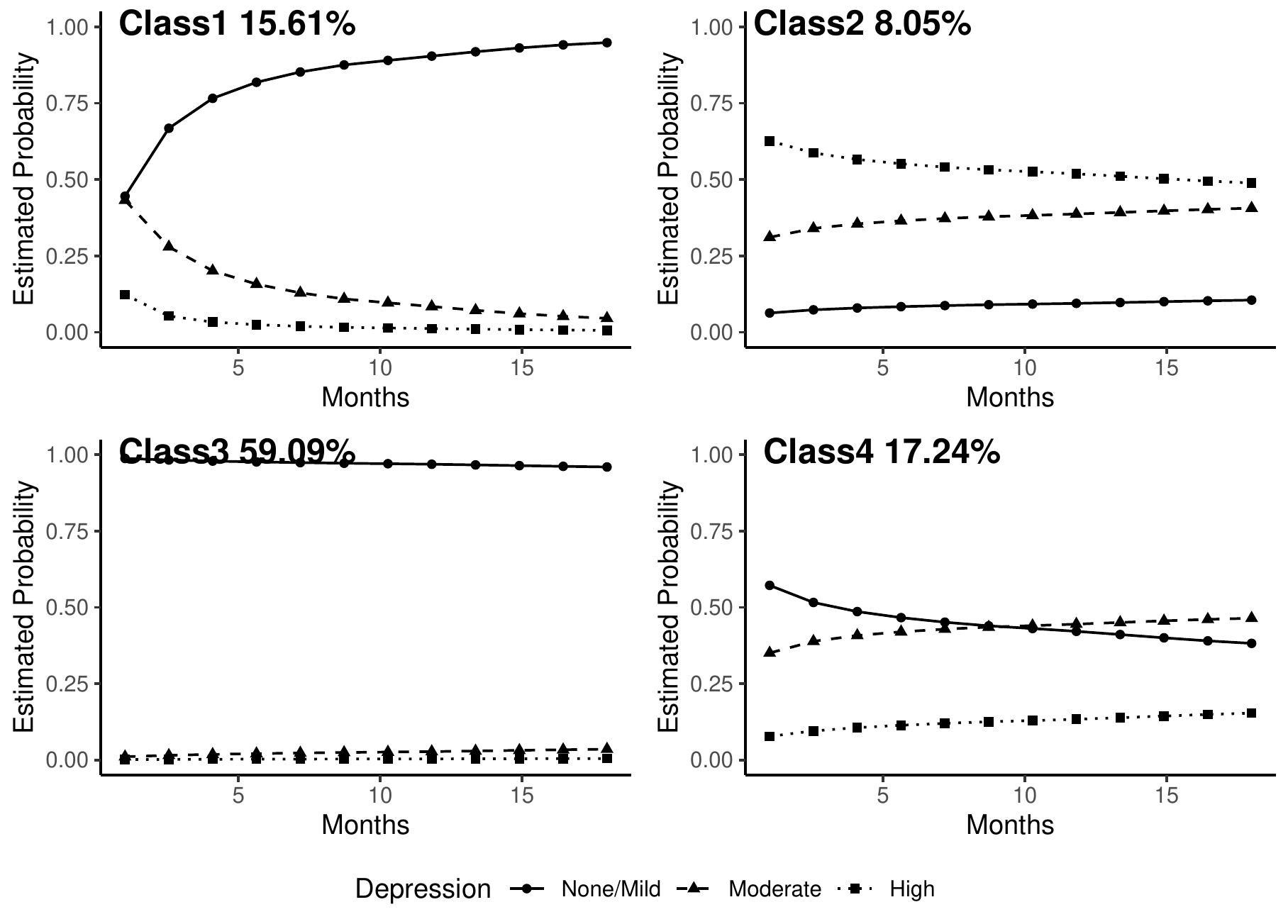}
    \caption{Estimated probabilities of each latent group found on DART ESAS-r Depression servery. Four distinct classes are found. Depression scores are grouped into None/Mild(0-3), Moderate(4-6) and High(7-10). Each line represents the probability for a patient in a given latent class falling into one of the depression categories over time.}
    \label{fig:depression_prob}
\end{figure}
\clearpage
\begin{table}[htb]
\resizebox{\textwidth}{!}{%
\begin{tabular}{|c|c|c|c|}
\hline
                                                          & Mplus (LCGA)           & SAS Traj               & R lcmm                 \\ \hline
Ordinal Data Model  & Censored Normal(one-side only)/Generalized probit & Censored Normal                          & Generalized probit                         \\ \hline
Random Effect                                             & Possible               & No                     & Possible               \\ \hline
Covariates                                                & Fixed and Time-varying & Fixed and Time-varying & Fixed and Time-varying \\ \hline
Order of Covariates       & {Unlimited in Principle}        & {Up to 3}              & {Unlimited in Principle} \\ \hline
MLE Method & Expectation-Maximization                          & {General Quasi-Newton} & Iterative Marquardt algorithm             \\ \hline
\multicolumn{1}{|l|}{Conditional Independence Assumption} & Flexible               & Yes                    & Flexible               \\ \hline
\end{tabular}%
}
\caption{Summary of main assumptions made in each implementation}
\label{tab:summary}
\end{table}

\clearpage
\begin{table}
\centering
\resizebox{\textwidth}{!}{%
\begin{tabular}{llllll}
\hline
Model                              & BIC of nclass =2                & BIC of nclass=3                            & 2$\Delta$BIC\tablefootnote{ If 2$\Delta$BIC $>$10, then it is a strong evidence to against simpler model (i.e. smaller groups).} & BIC of nclass=4                       & 2$\Delta$BIC \\ \hline
\textbf{Without Treatment Effect}  &                                 &                                            &              &                                       &              \\ 
\textit{Original Scale}            &                                 &                                            &              &                                       &              \\
R lcmm without covariates\tablefootnote{ Model without covariates captures the situation that although there is some underlying process to simulate the ESAS-r, we do not have the covariate information.}        & 155730.00                       & {\textbf{155683.00}} & 94           & 155798.00                             & -230         \\
R lcmm with  full covariates\tablefootnote{ Model with full covariates means all the covariates that used to simulate the ESAS-r score are included in the model. }       & 152604.00                       & {\textbf{151659.80}} & 1888.4       & 152027.00                             & -734.4       \\
SAS proc traj without covariates   & 86905                           & {\textbf{86806}}      & 198          & 86827                             & -42          \\
SAS proc traj with full covariates & 85377                           & 84913                                      & 928          & {\textbf{84822}} & 182          \\
\textit{Transformed to categories} &                                 &                                            &              &                                       &              \\
R lcmm without covariates          & 137847.1                        & {\textbf{137330.8}}  & 1032.6       & 137894.6                              & -1127.6      \\
R lcmm with  full covariates       & 136208                          & {\textbf{135665}}    & 1086         & 136232                                & -1134        \\
Mplus without covariates           & 18405.85                        & {\textbf{18401.04}}  & 9.62         & 18421                                 & -39.92       \\
Mplus with full covariates         & 15822.33                        & {\textbf{15148.45}}  & 1347.76      & 15189.01                              & -81.12       \\ \hline
\textbf{With Treatment Effect }     &                                 &                                            &              &                                       &              \\
\textit{Original Scale}            & {}         &                                            &              &                                       &              \\
R lcmm without covariates          & {250749}   & {\textbf{249902}}              & 847          & 250937                                & -1035        \\
R lcmm with  full covariates       & {\textbf{250024}}   & 250790                                     & -766         &                                       &              \\
SAS proc traj without covariates   & 133073                          & {132981.2}            & 91.8         & {\textbf{132979}}                               & 4.4          \\
SAS proc traj with full covariates & 132782.2                        & 132595.6                                   & 186.6        & {\textbf{132568.6}}       & 54           \\
\textit{Transformed to categories} &                                 &                                            &              &                                       &              \\
R lcmm without covariates          & 157941.5 & {\textbf{136825.5}}                                  &   42232      &              138843.6                       &       -4036.2       \\
R lcmm with  full covariates       & {166246.8} & 135893.7                                  & 60706.2        &                 {\textbf{129138.7}}                      &     13510         \\
Mplus without covariates           & {\textbf{91419.35}} & {91425.58}            & -12.46        &                                       &              \\
Mplus with full covariates         & 90837.34                        & {\textbf{90760.03}}            & 77.31        & 90796.98                              & -36.95    \\  \hline
\end{tabular}%
}
\caption{Summary of average BICs from each model with lowest BIC highlighted and logged Bayes factor (2$\Delta$BIC) computed for two consecutive class.}
\label{tab:BIC}
\end{table}
\clearpage
\begin{table}
\centering
\resizebox{\textwidth}{!}{%
\begin{tabular}{lccc}
\hline
                                      & Constantly low group(n=3000) & Increasing group(n=1000) & \multicolumn{1}{l}{Decreasing group(n=1000)} \\ \hline
\textbf{Without Treatment Effect}     & \multicolumn{1}{l}{}         & \multicolumn{1}{l}{}     & \multicolumn{1}{l}{}                         \\ \cline{1-1}
\textit{Original Scale}               & \multicolumn{1}{l}{}         & \multicolumn{1}{l}{}     & \multicolumn{1}{l}{}                         \\
R lcmm without covariates             & 95.03\%                      & 68.68\%                  & 71.70\%                                      \\
R lcmm with covariates                & 99.99\%                      & 98.04\%                  & 88.77\%                                      \\
SAS proc traj without covariates      & 99.32\%                      & 60.82\%                      & 93.17\%                                      \\
SAS proc traj with covariates         & 100\%                        & 97.36\%                  & 92.20\%                                      \\
\textit{Transformed to categories}    &                              &                          &                                              \\
R lcmm without covariates             & 92.15\%                      & 63.30\%                  & 93.20\%                                      \\
R lcmm with covariates                & 99.98\%                      & 99.46\%                  & 99.37\%                                      \\
Mplus without covariates              & 94.58\%                      & 71.43\%                  & 84.90\%                                      \\
Mplus with covariates                 & 99.99\%                      & 99.99\%                  & 99.25\%                                      \\ \hline
\textbf{With Latent Treatment Effect} & \multicolumn{1}{l}{}         & \multicolumn{1}{l}{}     & \multicolumn{1}{l}{}                         \\ \cline{1-1}
\textit{Original Scale}               & \multicolumn{1}{l}{}         & \multicolumn{1}{l}{}     & \multicolumn{1}{l}{}                         \\
R lcmm without covariates             & 66.88\%                      & 75.38\%                  & 66.83\%                                      \\
R lcmm with covariates                & 84.78\%                      & 90.99\%                  & 90.53\%                                      \\
SAS proc traj without covariates      & 87.85\%                      & 89.10\%                  & 83.97\%                                      \\
SAS proc traj with covariates         & 88.43\%                      & 79.59\%                  & 83.19\%                                      \\
\textit{Transformed to categories}    & \multicolumn{1}{l}{}         & \multicolumn{1}{l}{}     & \multicolumn{1}{l}{}                         \\
R lcmm without covariates             & 83.29\%                      & 83.12\%                  & 83.56\%                                      \\
R lcmm with covariates                & 92.97\%                      & 95.74\%                  & 94.56\%                                      \\
Mplus without covariates              & 87.83\%                      & 87.86\%                  & 85.62\%                                      \\
Mplus with covariates                 & 98.51\%                      & 97.16\%                  & 94.32\% \\ \hline                                    
\end{tabular}%
}
\caption{Summary of average correct posterior probability from each model}
\label{tab:pprob}
\end{table}

\clearpage
\begin{table}
\centering
\begin{subtable}{\linewidth}
\centering
\caption{SAS Proc Traj CNORM Convergence Information}
\begin{tabular}{lllll}
\hline
                     & nclass & BIC (N= 121380) & AIC       & log-likelihood \\ \hline
\textbf{Anxiety}     & 1      & -235259.1       & -235234.8 & -235229.8      \\
                     & 2      & -210711.8       & -210624.4 & -210606.4      \\
                     & 3      & -201600.8       & -201450.3 & -201419.3      \\
                     & 4      & -199141.6       & -198928.0 & -198884.0      \\
                     & 5      & \multicolumn{3}{c}{False Convergence}        \\ \hline
\textbf{Depression}  & 1      & -209618.5       & -209594.3 & -209589.3      \\
                     & 2      & -183295.1       & -183207.7 & -183189.7      \\
                     & 3      & -175502.6       & -175352.1 & -175321.1      \\
                     & 4      & -173351.5       & -173137.9 & -173093.9      \\
                     & 5      & -172414.5       & -172137.9 & -172080.9      \\
                     & 6      & -171759.8       & -171420.1 & -171350.1      \\
\multicolumn{1}{l}{} & 7      & \multicolumn{3}{c}{False Convergence}        \\ \hline
\end{tabular}
\end{subtable}
\vspace{0.5cm}

\begin{subtable}{\linewidth}
\centering
\caption{Mplus Threshold Convergence Information}\label{tab:second}
\begin{tabular}{lllll}
\hline
                    & nclass & BIC (N= 121380)       & AIC        & log-likelihood \\ \hline
\textbf{Anxiety}    & 1      &            &            &                                    \\
                    & 2      & 104916.412 & 104803.401 & -52387.701                         \\
                    & 3      & 100407.13  & 100205.38  & -50077.66                          \\
                    & 4      & 98903.59 2 & 98613.10   & -49270.50                          \\
                    & \textbf 5     & \textbf{97900.61}   & \textbf{97521.41}   & \textbf{-48713.61}                         \\
                    & 6      & 98011.40   & 97543.51   & -48713.61                          \\ \hline
\textbf{Depression} & 1      &            &            &                                    \\
                    & 2      & 126520.4   & 126407.4   & -63189.71                          \\
                    & 3      & 121425.7   & 121223.9   & -60586.94                          \\
                    & \textbf 4      & \textbf{119869.7}   & \textbf{119579.1}   & \textbf{-59753.57}                          \\
                    & 5      & 119892.7   & 119513.3   & -59709.64                          \\ \hline
\end{tabular}
\end{subtable}
\caption{Convergence Information for DART Anxiety and Depression using (A) SAS and (B) Mplus.}
\label{tab:convergence}
\end{table}
\clearpage
\appendix
\setcounter{table}{0}
\renewcommand{\thetable}{S\arabic{table}}
\setcounter{figure}{0}
\renewcommand{\thefigure}{S\arabic{figure}}
\section*{Appendix}
\begin{figure}[h]
    \centering
    \includegraphics[width=0.45\linewidth]{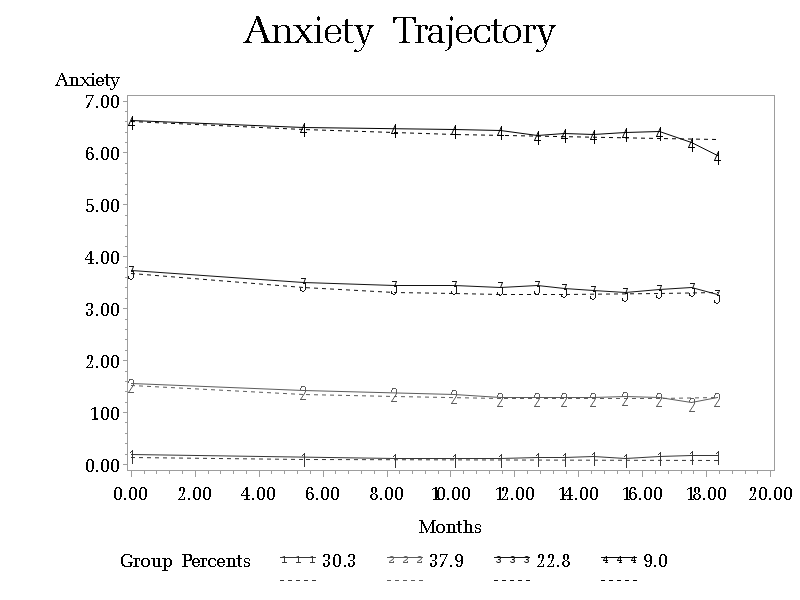}
    \includegraphics[width=0.45\linewidth]{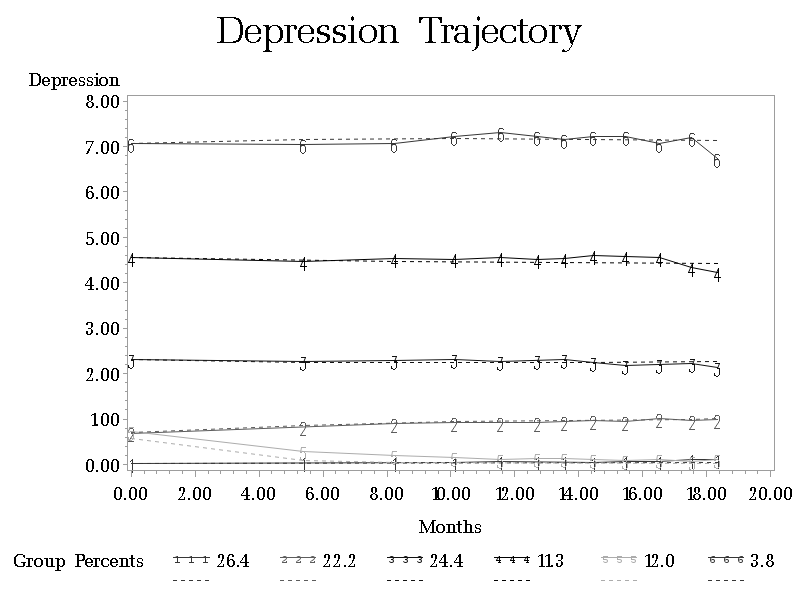}
    \caption{Trajectory plot of DART Anxiety and Depression using SAS Proc Traj. }
    \label{fig:sas_traj}
\end{figure}
\clearpage
\begin{table}
\centering
\resizebox{\textwidth}{!}{%
\begin{tabular}{|c|c|c|c|}
\hline
\multicolumn{4}{|c|}{Average  Computational  Time in Mins}                    \\ \hline
\multicolumn{2}{|c|}{Original Scale} & \multicolumn{2}{c|}{Categorical Scale} \\ \hline
R lcmm        & SAS Proc Traj        & R  lcmm catgeories        & Mplus      \\ \hline
0.20          & 0.00                 & 0.01                      & 0.00       \\ \hline
9.48          & 0.08                 & 4.93                      & 0.00       \\ \hline
55.15         & 0.20                 & 12.2                      & 1.23       \\ \hline
116.4         & 0.31                 & 80.98                          & 1.97       \\ \hline
\end{tabular}%
}
\caption{Average Computational Time of a model fit based on 5,000 patients and 12 surveys each with 1,000 replicates.}
\label{tab:computational time}
\end{table}
\end{document}